\documentclass{aa}
\usepackage{epsf,times}
\begin{document}

\thesaurus{06(08.19.3; 08.12.1; 08.06.3; 08.23.3; 13.09.6)}

\authorrunning{S.\ Philipp et. al}
\titlerunning{IRAS 17393$-$3004}
\title{IRAS 17393$-$3004, a late-type supergiant surrounded by a dust shell}

\author{S.\ Philipp\inst{1},
        R.J.\ Tuffs\inst{2},
        P.G.\ Mezger\inst{1},
        R.\ Zylka\inst{3,1}}

\offprints{S.\ Philipp, Bonn}
\mail{sphilipp@mpifr-bonn.mpg.de}

\institute{Max-Planck-Institut f\"ur Radioastronomie, Auf dem H\"ugel 69,
D-53121 Bonn, Germany (sphilipp/rzylka@mpifr-bonn.mpg.de)
\and
Max-Planck-Institut f\"ur Kernphysik, Saupfercheckweg,
D-69117 Heidelberg, Germany (rjt@ruby.mpi-hd.mpg.de)
\and
Institut f\"ur Theoretische Astrophysik, Tiergartenstra{\ss}e 15,
D-69121 Heidelberg, Germany (zylka@ita.uni-heidelberg.de)
}

\date{Received 7 July 1999 ; accepted 21 July 1999}

\maketitle

\begin{abstract}

Infrared observations of IRAS$\,$17393$-$3004, including a full scan 
ISOSWS~\footnote{ISO is an ESA project with instruments funded by ESA Member
States (especially the PI countries: France, Germany, The Netherlands and the
United Kingdom) with the participation of ISAS and NASA.} are presented. The 
ISOSWS spectrum shows two prominent dust emission peaks at $\lambda 
10.4\mu$m and 17.5$\,\mu$m confirming that the source is a luminous star 
surrounded by a dust shell. The spectrum of the star, deduced from 
modeling of the radiative transfer through a spherical shell using 
the {\it DUSTY} code (Ivezi\'{c} et al. 1997), can be explained along with 
other known properties in terms of an M4 supergiant with OH, SiO and H$_2$O 
maser emission and a large mass loss of $\sim 10^{-4}\,{\rm M}_\odot\,{\rm 
yr}^{-1}$\,. 

\end{abstract}


\section{Introduction}

Although not prominent in the IRAS all sky survey due to confusion with 
diffuse dust emission, IRAS$\,$17393$-$3004 
(galactic coordinates $l = -1^\circ19'57''$, $b = -0^\circ02'38''$) has
special interest as the brightest discrete source in the galactic plane within 
$\pm\,$2.0$\,$degrees of the galactic center at infrared wavelengths shorter 
than ca. 5$\,\mu$m. Strip scans in the range $-$2.0$\,\le\,l\,\le\,+$2.0 
degrees, taken with the ISOPHOT instrument on ISO as part of 
a program to investigate the physical state and characteristics of the Nuclear
Bulge (NB) (Philipp et al. 1999; Mezger et al. 1999) show the 
source to be brighter even than Sgr$\,$A West by a factor of 1.6 when viewed 
through a 99$''$ aperture in broad band filters centred on 3.6$\,\mu$m.

\begin{table}
\small{
\caption{Designation and position of IRAS 17393$-$3004}
\label{tababbrev}
\begin{center}
\begin{tabular}{|c|c|c|c|}
\hline
Designation& $\alpha_{1950}$ & $\delta_{1950}$ & Ref.\\
\hline
IRAS 17393$-$3004 & $17^h39^m22.4^s$ & $-30^\circ04'20''$& $[$1$]$\\
AFGL 1977 & $17^h39^m22.9^s$ & $-30^\circ04'23.0''$ & $[$2$]$\\
IRC-30316 & $17^h39^m30^s$ & $-30^\circ04'54.0''$ & $[$3$]$\\
\hline
\end{tabular}
\label{designations}
\end{center}
$[$1$]$ IRAS Point Source Catalogue, 1985\\
$[$2$]$ Grasdalen et al. 1983\\
$[$3$]$ Neugebauer \& Leighton 1969\\ 
}
\end{table}

The designation of IRAS$\,$17393$-$3004 (the name we adopt for this letter)
in three NIR/MIR source
catalogues together with the corresponding source positions are given in Table
\ref{designations}. The source has been extensively observed since its
discovery in the objective-prism Two-Micron Sky Survey by Hanson \& Blanco 
(1975), who classified it as a star of {\it spectral type M1}, 
{\it peculiar} with {\it emission near 8600\,\AA}.
IRAS$\,$17393$-$3004 is seen towards the
HII region Sgr E, which is believed to be located in the NB (Liszt 1992). It
is also listed in the catalogue of OH/IR stars (te Lintel Hekkert 
et al. 1989) as source No.\,145, where it is suggested to be a giant 
or an AGB star with a large mass outflow (\.{M} $> 10^{-5}\,{\rm 
M}_\odot\,{\rm yr}^{-1}$\,, Olnon et al. 1981). It shows an unusually large 
($\Delta v \sim 53$\,km\,s$^{-1}$) peak separation in the OH spectrum 
indicative of a massive central star; the central velocity of the OH emission 
determined using the separation of the two OH lines is close to zero 
(te Lintel Hekkert et al. 1989). IRAS 17393$-$3004 also shows SiO and H$_2$O 
maser emission (Hall 1990; Haikala et al. 1994; Lewis et al. 1995 and 
Sevenster et al. 1997). These maser characteristics suggest the central star 
to be a supergiant. Grasdalen et al. (1983) investigated the 
source in the NIR/MIR broadband continuum. Volk \& Cohen (1989) 
present an IRAS-LRS spectrum of this source showing structure at MIR 
wavelengths. In the IRAS data base the source is classified as having a high
probability to be variable. The ISOGAL survey excluded the region around 
IRAS$\,$17393$-$3004 (priv. comm. A.\,Omont). Although covered by the 
MSX~\footnote{Midcourse Space Experiment, a Ballistic Missile defense Office 
satellite.} survey, no parameters for this source are available as yet
(priv. comm. S.\,Price). No optical counterpart is seen on the 
POSS, neither has X-ray emission been found (ROSAT data archive; 
priv. comm. Y.\,Sofue for ASCA). 
There is a probable radio counterpart to IRAS$\,$17393$-$3004, in the form of
source$\,$9 in Table$\,$3 of Liszt (1992), which has a flux density of
21$\,$mJy in a band of width 12.5$\,$MHz centred at 1616$\,$MHz, and is 
extended (15$'' \times 10''$ at p.a.$\,$13$^\circ\,$). No counterpart is seen
in the National Radio Astronomy Observatory Very Large Array Sky Survey 
(NVSS 1995) at $\lambda\,$20$\,$cm. This may either indicate
a variable radio continuum and/or the presence of the 1612$\,$MHz
OH line in Liszt's measurement.
Here we report on observations of IRAS$\,$17393$-$3004 made with the SWS 
(de$\,$Graauw et al. 1996) and ISOPHOT (Lemke et al. 1996) instruments on 
board ISO, together with ground-based NIR photometry.

\section{Observations}

\subsection {NIR and FIR Continuum observations}

Coverages of a $7'\times5.5'$ field centred on IRAS$\,$17393$-$3004 were 
made using the ISOPHOT-C 3 $\times$ 3 pixel 
detector array in two spectral passbands. To 
avoid complete saturation of the detector the C50 (bandpass 40 - 93\,$\mu$m) 
and C105 (bandpass 89 - 125\,$\mu$m) filters were used.
Since IRAS$\,$17393$-$3004, although bright, is 
barely discernable at $\lambda > 40\,\mu$m against the structured background 
emission from the galactic plane at the angular resolution of 45$''$, the P32
mode was used to give the best 
available sky sampling. The data were reduced with version 7.1 of the 
ISOPHOT Interactive analysis package (Gabriel et al. 1997), and 
further corrected for the transient behaviour of the detector. An unresolved 
source was detected at the position of  IRAS1793-3005 in the C50 filter with 
flux density $37 \pm 12$\,Jy at a reference wavelength of 61\,$\mu$m (after 
colour correction assuming a $\nu^{+4}$ spectrum). A (confusion limited) upper 
limit of 20\,Jy was obtained in the C105 filter. 

In addition, the source was observed with the ESO-MPG 2.2m telescope on 
La Silla in June 1998 using the IRAC2B camera in the J, H, and K bands. 
Applying the data reduction 
techniques described in Philipp et al. (1999), we derived the flux densities 
given in Table \ref{tabnirflux}. 

\begin{table}
\caption{NIR flux density measurements of IRAS 17393$-$3004}
\label{tabnirflux}
\begin{center}
\begin{tabular}{|c|c|c|c|}
\hline
 & \multicolumn{3}{|c|}{NIR Band}\\
 & J & H & K\\ 
\hline
$\lambda / \mu$m & 1.25 & 1.65 & 2.2\\
$S_{\nu} / {\rm Jy}$ & 12.4 & 51 & 74\\
\hline
\end{tabular}
\end{center}
\end{table}

\subsection {Spectral observations with ISO-SWS}

\begin{figure}
\setlength{\unitlength}{1mm}
\begin{picture}(55,63)
\includegraphics{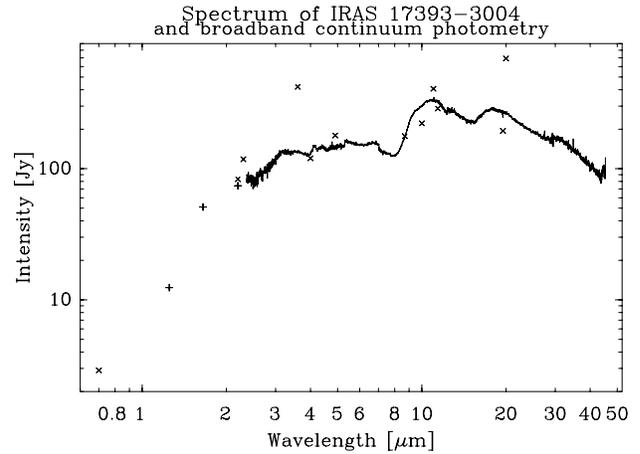}
\end{picture}
\caption{ISOSWS spectrum of IRAS 17393$-$3004 and broadband continuum
observations in the NIR and MIR ({\tt `x'} corresponds to measurements in
Grasdalen et al. (1983); {\tt `+'} to our NIR photometry).}
\label{figgrasdalen}
\end{figure}

As a follow-up to the detection of IRAS$\,$17393$-$3004 as such a prominent
source in the scans in $l$ with ISOPHOT, a full spectral scan was made on 
20$^{th}$ February 1998 in the 2.38$\,$-$\,$45$\mu$m spectral range using 
the SWS instrument in AOT01 mode with scanning speed 3. 
The pointing was specified directly towards the probable radio counterpart
measured by Liszt (1992) at (B1950) $17^h39^m22.7^s$ 
$-30^\circ04'16.0''$, thus ensuring that the full 
15$''\,$x$\,$10$''$ extent of the radio source (at p.a.$\,$13$^\circ\,$)
would be enclosed by the SWS apertures for the position angle of the 
observation. The spectrum 
was fully reduced with IA and OSIA (version of November 25$^{th}$ 1998 and 
Version 1.0 respectively; Thronley et al. 1997) 
starting with the latest pipeline data product (OLP V7.01). Deglitching, dark 
current subtraction, tail modeling, photometric checks as well as 
updown-corrections and defringing were applied. 
The calibration used standard tables within the IA derived from 
observations of Uranus; the overall uncertainty 
is estimated as $\sim 30\%$ (priv.comm. SWS data centre).
In Fig.\,\ref{figgrasdalen} 
the ISOSWS spectrum is shown supplemented by the earlier
broadband continuum observations from Grasdalen et al. (1983),
and by the recent NIR observations.

The SWS spectrum shows prominent broad emission and/or absorption features 
due to dust and molecules, superimposed on a smooth continuum. As confirmed by
the detailed modeling (Sect. 3.2) the continuum seen by SWS makes a
transition from reddened photospheric-dominated emission in the NIR
wavelengths to warm dust emission in the MIR.
The spectral peaks at 10.4 and 17.5$\,\mu$m are most readily identified
as silicate emission features. Although the shorter wavelength
feature is normally centred at shorter wavelengths in the 
9$\,$-$\,$10$\,\mu$m range, it appears here
superimposed on broad-band molecular 
absorption from gaseous SiO, whose absorption minimum is seen around 
8$\,\mu$m. SiO bandheads are seen in the range 4.0 
and 4.3\,$\mu$m (this structure is discussed for cool stars in detail by 
Aringer et al. 1999). The other very deep molecular absorption 
bands in the $\lambda$ 2.5 to 9$\,\mu$m region can be ascribed to H$_2$O.  
In this regime the spectrum is very similar to synthetic spectra of K and M 
giants and supergiants (Decin et al. 1997; Tsuji et al. 1997); comparison
shows that specifically H$_2$O is prominent in IRAS 17393$-$3004. 
Taken together, the solid-state and molecular features indicate that 
IRAS 17393$-$3004 is an oxygen-rich high mass-loss star. Although 
[Si\,II] is seen at $\lambda$ 34.8$\,\mu$m,
other fine structure lines are not prominent, the feature 
at $\lambda$ 11.05$\,\mu$m being an artifact of the calibration (priv. comm. 
D.\,Lutz and D.\,Kunze).

\subsection {Infrared Variability?}

Although the IRAS variability flag was set, we were not able to
confirm variability between the IRAS and ISO epochs after a detailed
comparison of the SWS spectrum and HIRES IRAS maps.
Due to confusion, the source 
is not seen in IRAS HIRES maps in the 60 and 100\,$\mu$m bands and barely 
seen in the 25\,$\mu$m band. Both the flux density obtained from the 
12$\,\mu$m HIRES map (290$\,\pm\,$30$\,$Jy) and the value 256$\,$Jy derived by 
Haikala et al. (1994) and IRAS Point Source Catalogue (1985) and for the same 
band are consistent with a spectral average of the SWS data over the spectral 
response function of IRAS through the 12 micron band, which yields 281 Jy. 

\begin{figure}
\setlength{\unitlength}{1mm}
\begin{picture}(55,61)
\includegraphics{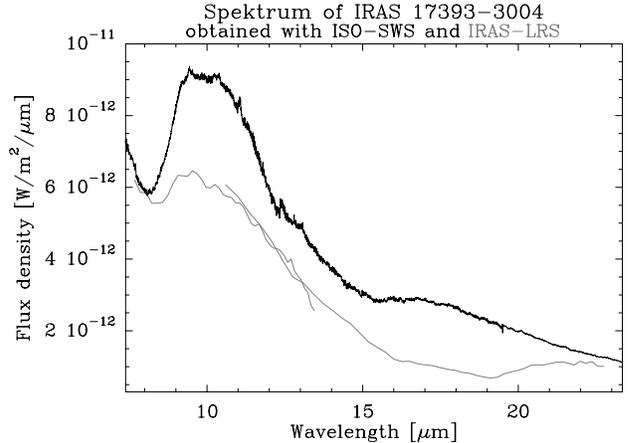}
\end{picture}
\caption{Part of the ISOSWS spectrum of IRAS 17393$-$3004 (heavy curve)
compared to the IRAS-LRS spectrum (dotted curve; Volk \& Cohen 1989).}
\label{figlrssws}
\end{figure}

Some evidence for spectral variability is however provided by 
a comparison of the IRAS-LRS spectrum 
(Volk \& Cohen 1989) and the SWS spectrum (Fig.\,\ref{figlrssws}).
While the overall spectral distribution is similar, the line to continuum
contrast of the silicate features differ between the datasets. The continuum 
level on the IRAS-LRS spectrum lies markedly below
that of the SWS level, though it is not clear why this should be bearing in
mind the consistency of the IRAS survey data with the SWS photometry and the 
fact that the Haikala et al. (1994) IRAS 12$\,\mu$m flux density was used
to calibrate the LRS spectra, for which there were no obvious technical
problems (priv. comm. R.\,Waters \& LRS database 1987). Probably these
differences might also be caused by the extraction of the spectrum from a 
crowded region, showing a strong gradient in the background (K.\,Volk, priv. 
comm). Inspection of Fig. 1 shows that the Grasdalen et al. (1983) flux 
densities in the 2$\,$-$\,$20$\,\mu$m spectral range, measured in the 1970's, 
may also support a picture of spectral variability. While most of the data 
points are consistent with the SWS data, the values for the pass-bands centred
on $\lambda 3.6\,\mu$m and 20$\,\mu$m are higher by more than a factor 2. One 
can speculate that the 
20$\,\mu$m observations may be affected by atmospheric variations;
the discrepancy of the $3.6\,\mu$m observation is not really understood.

\section{Modeling of the spectrum}

\subsection {Nature of the star}

The previous observations listed in Sect.$\,$1 suggest that the
star powering IRAS$\,$17393$-$3004 is a 
supergiant of spectral type M. Elias et al. (1981) 
investigated late-type supergiants in our Galaxy and in the 
Magellanic Clouds, obtaining $M_{\rm V} =-8$ and $M_{\rm K}=-12$
as upper limits. The extreme case corresponding to these
limits is for spectral type M4I (see e.g. Lang 1992), for which we 
estimate a stellar luminosity of $\sim\,$1.3$\times10^5\,{\rm L_\odot}$, 
using the definition for absolute bolometic luminosity given by Lang (1974, 
Eq.\,(5)-(237)). Combining this luminosity with the observed K band flux density
of $S_{\rm K}^\prime = 74$\,Jy (2.33\,mag) yields an upper limit for the 
distance $D \le 7.335$\,kpc (disregarding any extinction by dust), placing 
IRAS 17393$-$3004 in front of the NB. This is also supported by the system 
velocity of the maser of $\sim -7$\,km\,s$^{-1}$ (Haikala et al. 1994), 
which does not coincide with the rotation curve of the NB. 
In fact the near-IR colours are consistent with the exciting star being 
close to the limiting case of an M4I supergiant. For such a star we 
estimate, based on the fact that IRAS$\,$17393$-$3004 has no optical 
counterpart on the POSS, that the sum of dust 
extinction in the shell (expressed in terms of visual extinction) 
$A_{\rm vis,shell}$ and in the intervening dust between the sun and the 
supergiant $A_{\rm vis,ISM}$ would be 
$A_{\rm vis,shell} + A_{\rm vis,ISM} \ge 14$\,mag. Such an extinction 
would place our NIR flux densities in the correct regime of colours given by 
Elias et al. (1985) in Table\,10 for M4\,I stars. 

\subsection {Modeling the Dust Emission }

The above considerations have led to a scenario for IRAS$\,$17393$-$3004
in which a supergiant
of spectral type M4 undergoes a strong mass outflow resulting in a shell
of gas and dust surrounding the star for which we estimate a total
luminosity close to $ \sim 1.3\times10^5\,{\rm L_\odot}$. 
The inner radius of the dust shell R$_{\rm in}$ is constrained by the fact that
grains are probably needed to form a sufficiently copious amount of
molecules to account for the maser source. For a star with the above 
luminosity masing must occur within $\sim\,$10$^{16}$\,cm for 
OH masing (Engels et al. 1983) and $\le 10^{15}$\,cm for 
H$_2$O masing (Elitzur 1992). The outer bound on the 
masing region for H$_2$O is indeed quite close 
to the sublimation radius for black body grains of 
$\sim\,$3$\,$10$^{14}$\,cm for this stellar
luminosity and a sublimation temperature of 1230$\,$K.

To check that the SWS spectrum
of the dust emission is consistent with this scenario, and to derive
parameters for the dust outflow,
we have modeled the dust emission using the 
{\it DUSTY} code (Ivezi\'{c} et al. 1997; Ivezi\'{c} \& 
Elitzur 1997 and 1995). This solves the spherical radiative transfer problem 
employing a scaling approach, taking the grains to be distributed from 
a condensation radius to at least the radius beyond which the grains become too
cool to contribute to the SWS spectrum. In fact it is also possible to
successfully model the dust emission spectrum in terms of a completely
optically thin shell with a much larger inner boundary and an outer radius
cutoff. If true this might imply some variability in the recent dust 
production rate. However we regard this as inconsistent with the evidence 
(from the H$_2$O maser emission as discussed above) that grains are being 
formed at or near the condensation radius. Moreover, there is no completely 
convincing evidence for variability in the warm dust emission 
(see Sect.\,2.3).

\begin{figure}
\setlength{\unitlength}{1mm}
\begin{picture}(55,62)
\includegraphics{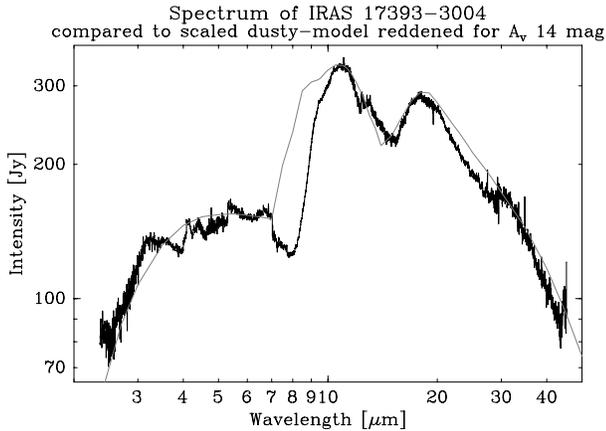}
\end{picture}
\caption{SWS spectrum of IRAS*$\,$17393$-$3004 overlaid with
the best fit {\it DUSTY} model (see text).}
\label{figmodel}
\end{figure}

{\it DUSTY}'s output provides a self-similar set of solutions possessing  
scaling properties (Ivezi\'{c} \& Elitzur 1997). This allows a modeling in 
two steps. In a first step we adopt $\tau_{\rm K} \sim 1.57$ (corresponding to 
$A_{\rm vis} \sim 14$\,mag) as the optical depth of the dust shell, 
$T_{\rm BB*} \sim 3000$\,K as the stellar temperature and 1230$\,$K as
the condensation temperature. Subsequently, we fine tune
the observed spectrum by varying the temperature of the illuminating 
black-body source, and the condensation temperature.
Variations of $\sim 10\%$ in these parameters are clearly noticeable in the 
shape of the model spectrum. No attempt is made to model the molecular and
atomic absorption/emission line spectrum.
For the density profile we used the analytical approach for a 
radiatively driven wind, available in {\it DUSTY} (see Ivezi\'{c} et al. 
1997). The chemical composition of the dust was also varied to find the best 
fitting model using the six common grain types contained in {\it DUSTY} 
and the standard grain size distribution from Mathis et al. (1977). 
The gas to dust ratio is set to 200 by {\it DUSTY}.  

The best fit to the ISOSWS spectrum is shown in Fig.\,\ref{figmodel}. It was
obtained for a temperature of the embedded source of 
$T_{\rm BB*}\sim$2980$\,$K, $A_{\rm vis}\sim$14$\,$mag, and a dust 
condensation temperature of 1230\,K. 
The dust composition was set to 60\% silicates and 40\% graphite 
grains within the model of Draine \&  Lee (1984).  The fraction of graphite
grains might be quite high for an oxygen-rich star, but such a high 
fraction is necessary to fit the observed  flat slope at longer wavelengths. 
From discussions with M.\,Elitzur (priv. comm.) we learned, that some 
asymmetrical silicate 
grains show properties very similar to the graphite grains used here, and that 
this kind of grains are actually found in oxygen-rich stars. In fact such 
grains will be included in the subsequent version of {\it DUSTY}.

The largest difference between the observed and modeled spectra occurs in the 
range $\lambda 7 - 10 \mu$m. We attribute this partly to the presence of SiO 
absorption which is not modeled by {\it DUSTY}. In addition this feature 
may be partly due to the presence of some uncommon silicates in the shell of
IRAS 17393$-$3004, in which case the shape of the spectral feature would not be
typical. In addition there could be a combination of emission by the dust shell
and absorption by the interstellar dust. This suggestion is supported by the 
fact that the ISM silicate absorption peaks at a slightly shorter wavelength 
than the usual 9.7 micron emission peak. 
In addition the observed spectrum shows a weak emission feature at about
$\lambda 30\,\mu$m. This may either indicate emission from cooler dust in a 
detached shell, or dust emission features from crystalline Mg-rich silicates
in the 20$\,$-$\,$45$\,\mu$m range (e.g. Henning 1999). 
The other smaller variations between the model fit and the 
observed spectrum presumably reflect possible variations of the chemical 
composition and/or grain size.

In a second step we scale the model by integrating the flux density of the
model spectrum $F_{\lambda}$ to find the bolometric luminosity, which is
normalized to $L_* \sim 1.3\times10^5\,{\rm L_\odot}$, as estimated in 
Sect.\,3.1. This yields a distance of the star of $D = 4.73$\,kpc if all 
extinction occurs in the shell, consistent with the crude estimate $D \le 
7.335$\,kpc given in Sect.\,3.1. Clearly, some line of sight extinction
is to be expected, which would reduce the intrinsic absorption in the shell
from the value adopted in the modeling, and reduce the distance to the star.
As discussed below, this is not a very sensitive parameter, especially in 
comparison with the condensation temperature parameter.
To explore this effect we also made calculations assuming a line of sight 
extinction of 5\,mag. This would reduce the extinction within the shell to 
$\sim 9$\,mag. Our computations show
that this model will lead to a condensation radius which is only $\sim 10\%$ 
smaller and to a mass loss rate which is only $\sim 20\%$ lower. This leads 
us to consider our results to be upper limits of the actual stellar and 
associated parameters with an estimated uncertainty of  $\sim 30\%$, mainly 
caused by the calibrational uncertainty of the ISOSWS data.

We return to our original model with all extinction due to dust in the shell 
surrounding the central star. In this case the condensation radius scales 
to $\sim 7.8\times10^{14}$\,cm for the silicate grains, consistent with the 
maximum radius of $\sim\,$10$^{15}$\,cm for the H$_2$O maser emission 
(Elitzur 1992). The derived values for the gas mass loss rate and velocity 
of the wind are $\sim$1.0$^{-4}\,{\rm M}_\odot\,{\rm yr}^{-1}$ and 
$\sim\,$18.7\,km\,s$^{-1}$ respectively. The model also yields a 
mass of $M_* \le 27.3\,{\rm M}_\odot$ for the embedded star. All these 
parameters are in remarkably good agreement with the 
basic observational constraints discussed in 
Sect.\,3.1, in particular the identification of the star as a type M4I
supergiant.

\section{Summary}

Combining ground- and spaceborne observations we have obtained
the spectral energy distribution for the IR/OH star
IRAS$\,$17393$-$3004 in the wavelength 
range $1.25 \le \lambda / \mu$m$ \le 60$. The near-IR spectrum
has JHK colours corresponding to an M4 star reddened by dust with
$A_{\rm vis}\,\sim\,$14$\,$mag. The MIR spectrum shows prominent 
silicate peaks at 10.4 and 17.5$\,\mu$m as well as an SiO absorption
around 8$\,\mu$m and SiO bandheads in the 4.0 - 4.3\,$\mu$ range. Taken
together, the solid-state and molecular features confirm the picture
already suggested from the SiO, OH and H$_2$O masing activity that 
IRAS 17393$-$3004 is a luminous oxygen-rich star with a high mass-loss 
dusty wind. Neglecting line of sight extinction,
the broad band spectral energy distribution of the dust emission
and the extinction in the near-IR can be fitted by a spherically 
symmetrical outflow of a silicate and graphite grain mixture 
illuminated by an M4I supergiant situated at a distance of $\le$ 4.73$\,$kpc
with luminosity $\sim 1.3\times10^5\,{\rm L_\odot}$, dust production
rate 5$\,$10$^{-7}$M$_{\odot}\,{\rm yr}^{-1}$ and wind velocity
$\sim\,$18.7\,km\,s$^{-1}$.
For a distance of 4.73$\,$kpc, the shell's inner boundary 
diameter corresponds to an angular size of $\sim 0.022''$, which will make it 
accessible in future to quantitative observations with the VLA (maser emission)
and VLTI (MIR dust emission) respectively.

\vspace*{1mm}
\noindent {\it Acknowledgements.}
We gratefully acknowledge the support provided by the ISO data centres at MPIA 
in Heidelberg and MPE in Garching. We benefited much from discussions with 
M.\,Elitzur, H.-P.\,Gail, Th.\,Henning, H.-U.\,K\"aufl, D.\,Kunze, D.\,Lutz, 
K.M.\,Menten, R.\,Waters and A.A.\,Zijlstra. Our special thanks go to the 
referee K.\,Volk for his comments, from which the paper profited much.


\end{document}